\documentclass[journal]{IEEEtran}
\usepackage[utf8]{inputenc}

\usepackage[hidelinks]{hyperref}
\usepackage{cite}
\usepackage{microtype}
\usepackage[nolist]{acronym}
\usepackage{dirtytalk}
\usepackage{pifont}
\usepackage{gensymb}
\usepackage{orcidlink}
\usepackage[subtle]{savetrees}

\usepackage{booktabs, dcolumn}
\usepackage{multirow}

\usepackage{bm}
\usepackage{mathtools}
\usepackage{physics, amsmath,amssymb,amsfonts}

\usepackage{caption}
\usepackage{subcaption}
\usepackage[export]{adjustbox}
\usepackage{graphicx}
\usepackage{pgfplots}
\DeclareUnicodeCharacter{2212}{−}
\usepgfplotslibrary{groupplots,dateplot}
\usetikzlibrary{quotes,arrows.meta, calc, patterns.meta, shapes.multipart, shapes.arrows, shapes.misc, spy, decorations.pathreplacing, arrows.meta, backgrounds, shapes.geometric, intersections, matrix}
\pgfplotsset{compat=1.15}

\def\BibTeX{{\rm B\kern-.05em{\sc i\kern-.025em b}\kern-.08em
    T\kern-.1667em\lower.7ex\hbox{E}\kern-.125emX}}

\begin{acronym}
\acro{DFT}{discrete Fourier transform}
\acro{MVDR}{minimum variance distortionless response}
\acro{PDF}{probability density function}
\acro{MMSE}{minimum mean square error}
\acro{ML}{maximum likelihood}
\acro{SNR}{signal-to-noise ratio}
\acro{MAP}{maximum a posteriori}
\acro{ASR}{automatic speech recognition}
\acro{POLQA}{perceptual objective listening quality analysis}
\acro{MOS}{mean opinion score}
\acro{PESQ}{perceptual evaluation of speech quality}
\acro{EM}{expectation maximization}
\acro{DNN}{deep neural network}
\acro{LSTM}{long short-term memory}
\acro{FF}{feed-forward}
\acro{cIRM}{complex ideal ratio mask}
\acro{IRM}{ideal ratio mask}
\acro{RIR}{room impulse response}
\acro{ATF}{acoustic transfer function}
\acro{CRNN}{convolutional recurrent neural network}
\acro{CNN}{convolutional neural network}
\acro{STFT}{short-time Fourier transform}
\acro{CQS}{continuous quality scale}
\end{acronym}

\def\vec#1{\ensuremath{\mathbf{#1}}}

\DeclareMathOperator*{\argmin}{arg~min~}

\newcommand{\vecY}{\vec{Y}}
\newcommand{\vecV}{\vec{V}}

\newcommand{\vecd}{\vec{d}}
\newcommand{\vech}{\vec{h}}
\newcommand{\vecPhi}{\vec{\Phi}}

\newcommand{\numMics}{\ensuremath{C}}
\newcommand{\numFreqs}{\ensuremath{F}}
\newcommand{\numTimes}{\ensuremath{T}}
\newcommand{\micidx}{\ensuremath{\ell}}
\newcommand{\sampleidx}{\ensuremath{t}}
\newcommand{\freqbinidx}{\ensuremath{k}}
\newcommand{\timeframeidx}{\ensuremath{i}}

\begin{document}
\title{Insights Into Deep Non-linear Filters for Improved Multi-channel Speech Enhancement}

\author{Kristina~Tesch\,{\orcidlink{0000-0002-6458-8128}},~\IEEEmembership{Student Member,~IEEE}, and
        Timo~Gerkmann\,{\orcidlink{0000-0002-8678-4699}},~\IEEEmembership{Senior Member,~IEEE}%
\thanks{The authors are with the Signal Processing Group, Department of Informatics, Universität Hamburg, 22527 Hamburg, Germany (e-mail: kristina.tesch@uni-hamburg.de; timo.gerkmann@uni-hamburg.de).}%
}

\maketitle

\begin{abstract}
The key advantage of using multiple microphones for speech enhancement is that spatial filtering can be used to complement the tempo-spectral processing. In a traditional setting, linear spatial filtering (beamforming) and single-channel post-filtering are commonly performed separately. In contrast, there is a trend towards employing \acp{DNN} to learn a joint spatial and tempo-spectral non-linear filter, which means that the restriction of a linear processing model and that of a separate processing of spatial and tempo-spectral information can potentially be overcome. 
However, the internal mechanisms that lead to good performance of such data-driven filters for multi-channel speech enhancement are not well understood.

Therefore, in this work, we analyse the properties of a non-linear spatial filter realized by a \ac{DNN} as well as its interdependency with temporal and spectral processing by carefully controlling the information sources (spatial, spectral, and temporal) available to the network. We confirm the superiority of a non-linear spatial processing model, which outperforms an oracle linear spatial filter in a challenging speaker extraction scenario for a low number of microphones by 0.24 POLQA score. Our analyses reveal that in particular spectral information should be processed jointly with spatial information as this increases the spatial selectivity of the filter. Our systematic evaluation then leads to a simple network architecture, that outperforms state-of-the-art network architectures on a speaker extraction task by 0.22 POLQA score and by 0.32 POLQA score on the CHiME3 data.
\end{abstract}


\section{Introduction}
In our everyday life, speech understanding often takes place in noisy environments. This can be, for example, a conversation in a crowded restaurant, a phone call in a busy train station or the use of a voice control system in a driving car. To enable devices such as hearing aids or voice-controlled assistants to function in these challenging acoustic environments, speech enhancement algorithms are employed to improve the speech quality and intelligibility of the target speech signal. 

Traditionally, many algorithms utilized a \ac{STFT} signal representation and derived an analytical clean speech estimator from a statistical model, e.g., \cite{1984ephraimSpeechEnhancementUsing, lotter2005SpeechEnhancementMAP, Erkelens2007MinimumME, becker2016}. While this has led to many interpretable and computationally lightweight algorithms, the derivations often require restricting and simplifying assumptions, e.g., independent time-frequency bins, to keep the problem tractable. This is in contrast to \ac{DNN}-based algorithms, which do not need an explicit model, but learn to recognize complex dependencies directly from training data. In the domain of single-channel speech enhancement, these DNN-based algorithms, have been dominating the state of the art for a couple of years now, e.g.,  \cite{tan2018convolutional, giri2019waveUnet, koizumi2021sepformer, Hao2021fullsubnet}. 

\begin{figure}
    \begin{minipage}[]{0.1\linewidth}
        \subcaption{}\label{fig:1-separated}
    \end{minipage}
    \begin{adjustbox}{minipage=0.85\linewidth}
        \centering
        \includegraphics{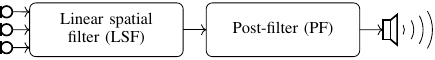}
    \end{adjustbox}
    
    \begin{minipage}[]{0.1\linewidth}
        \subcaption{}\label{fig:1-jointfilter}
    \end{minipage}
    \begin{adjustbox}{minipage=0.85\linewidth}
        \centering
        \includegraphics{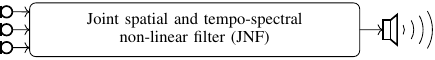}
    \end{adjustbox}
    
    \begin{minipage}[]{0.1\linewidth}
        \subcaption{}\label{fig:1-separated2}
    \end{minipage}
    \begin{adjustbox}{minipage=0.85\linewidth}
        \centering
        \includegraphics{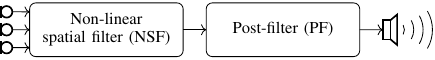}
    \end{adjustbox}
    \caption{(a) The traditional two-step processing using a linear spatial filter (beamformer) followed by a single-channel postfilter. (b) A joint spatial and tempo-spectral non-linear processing scheme that we implement using \acp{DNN} in this work. (c) Two-step processing scheme, however, not only the postfilter performs non-linear filtering but also the spatial filter.}
    \label{fig:1-comparison}
\end{figure}

While single-channel speech enhancement approaches exploit tempo-spectral signal characteristics to perform the enhancement, multi-channel approaches can additionally leverage spatial information by using multiple microphones. Commonly, this is done by employing a linear spatial filter, a so-called beamformer. Figure \ref{fig:1-separated} illustrates a traditional multi-channel processing pipeline, which first applies the linear spatial filter and then adds a single-channel post-filter in a second step. The post-filter can be either linear or non-linear. In our prior work \cite{tesch2019NonlinearSpatialFiltering, tesch2021nonlinearspatialfilteringtasl}, we have demonstrated that separation into a linear spatial filter and a post-filter is generally not optimal in the \ac{MMSE} sense unless we restrict the noise distribution to be Gaussian. However, if a non-Gaussian distribution is assumed, the resulting analytical solution is overall non-linear and joins the spatial and spectral processing as illustrated in Figure \ref{fig:1-jointfilter}. Our experimental evaluation in \cite{tesch2021nonlinearspatialfilteringtasl} has shown great potential for a joint non-linear spatial-spectral filter, but has also led to the conclusion that the estimation of required higher-order parameters limits the practical applicability of the analytic estimator. However, \acp{DNN} provide a data-driven way to implement practical joint spatial and tempo-spectral non-linear filters (JNF). 

A very influential paper on multi-channel speech enhancement using \acp{DNN} has been the paper by Heymann et al. \cite{heymann2015blstm}, who propose to use a \ac{DNN} for estimating the parameters of a linear spatial filter. Also others have proposed approaches along this line of research, e.g., \cite{togami2019itakura, 2018liuNeuralNetworkBased, xiao2016deepbeamforming}. However, using a \ac{DNN} for parameter estimation does not allow for a more general non-linear processing model nor does it permit the exploitation of interdependencies between spatial and tempo-spectral information during processing. In contrast, a variety of data-driven multi-channel filters have been proposed recently \cite{li2019narrowband, chakrabarty2019, tolooshams2020cadunet, wang2020complexspectralmapping, li2022eabnet, halimeh2022cospa}. These implicitly drop the linearity assumption and integrate spatial and tempo-spectral processing steps such that this class of joint non-linear approaches is fundamentally different and potentially more powerful than DNN-driven linear spatial filters, aka neural beamformers. Good performance as been reported for these deep non-linear filters, but the internal mechanisms that lead to good performance are not well understood. This, however, is essential for a deliberate design of a neural network architecture that fully unlocks the potential of neural networks for multi-channel speech enhancement. 

In this work, we investigate the internal functioning of DNN-based (joint) non-linear filters for multi-channel speech enhancement. To learn about the role of a non-linear spatial filter and the interdependency between spatial and tempo-spectral information, we consider a second separated approach, which combines a non-linear spatial filter with an independent post-filter. An illustration of this setup is given in Figure \ref{fig:1-separated2}. A systematic comparison of the three approaches outlined in Figure \ref{fig:1-comparison} then allows us to assess what makes for good spatial filtering performance: Is non-linear as opposed to linear spatial filtering the main factor for good performance? Or is it rather the interdependency between spatial and tempo-spectral processing? And do temporal and spectral information have the same impact on spatial filtering performance? 

This paper is based on our recent conference paper \cite{tesch2022interspeech}, but the experimental evaluation here goes far beyond the results presented previously. Specifically, we propose new experimental designs to investigate the spatial filtering performance of a DNN-based joint filter. This then allows for a discussion of the spatial selectivity of the different approaches. We include a comparison with state-of-the-art approaches, showing that the joint non-linear filter obtained by our systematic evaluation outperforms them and, furthermore, we extend our evaluations from the speaker extraction task to the CHiME3 dataset. The latter then enables us to assess the role of the dataset characteristics with respect to the previously mentioned research questions. 

In a recent study, also Tan et al. \cite{tan2022spatiospectralfilter} compare the performance of a joint spatial and tempo-spectral non-linear filter with a DNN-driven linear spatial filter plus additional post-filter (i.e. Figure \ref{fig:1-separated} versus Figure \ref{fig:1-jointfilter}). While they report comparable performance for these two approaches, in this paper, in line with our theoretical findings in \cite{tesch2021nonlinearspatialfilteringtasl}, we demonstrate the conceptual superiority of a joint non-linear spatial and tempo-spectral filter by outperforming an \emph{oracle} linear spatial filter plus post-filter. Furthermore, our work adds additional value beyond a general performance comparison of the two approaches by presenting experiments that allow for insights into the internal mechanisms underlying a well-performing joint non-linear filter.

The remainder of this paper is structured as follows. Section \ref{sec:background} introduces the signal model and provides a detailed overview of traditional and DNN-based spatial filtering. In Section \ref{sec:netarcs}, we introduce a set of DNN-based filter variants, which will be analyzed thoroughly to provide insights into the separability of spatial processing and post-filtering (Section \ref{subsec:evalseparability}) and the interdependency between spatial and tempo-spectral processing (Section \ref{subsec:evalinterdependency}). In Section \ref{sec:mixeval}, we provide a comparison with recent state-of-the-art methods and, in Section \ref{sec:chimeeval}, we report results for the CHiME3 dataset.

\section{Background and related work}\label{sec:background}
\subsection{Signal model}
We consider the task of extracting a single target speaker from a recording obtained in a noisy and reverberant environment. The noise signals may be environmental noise or concurrent speakers. The noisy mixture signals are captured by a microphone array with $\numMics$ microphone channels. In the time-domain, the speech signal uttered by the target speaker and recorded by the $\ell$th microphone can be written as the convolution of the non-reverberant speech signal $s(\sampleidx )$ and the \ac{RIR} $h_{\micidx} (\sampleidx)$ describing the propagation path between the speaker and the $\ell$th microphone \cite{doclo2015assistedlistening}:
\begin{equation}
    x_{\micidx} (\sampleidx) = s(\sampleidx) * h_{\micidx} (\sampleidx). 
\end{equation}
Note that, besides the room characteristics, $h_{\micidx} (\sampleidx)$ also allows to model the characteristics of the loudspeaker and the microphones.

We transform the time-domain signal $x_{\micidx}(t)$ into the frequency domain using a short-time Fourier transform (STFT) to obtain complex spectral coefficients $X_{\micidx}(\freqbinidx, \timeframeidx)\in \mathbb{C}$ with frequency-bin index $\freqbinidx$ and time-frame index $\timeframeidx$. Based on an additive signal model, the mixing process in the frequency domain is given by
\begin{IEEEeqnarray}{rCl} 
Y_{\micidx}(\freqbinidx,\timeframeidx)&=&X_\micidx(\freqbinidx, \timeframeidx) + V_\micidx(\freqbinidx,\timeframeidx).
\end{IEEEeqnarray}
with $V_\micidx(\freqbinidx,\timeframeidx)$ denoting the noise signal recorded at the $\micidx$th microphone. 
We use bold face symbols to refer to the vector stacking the STFT coefficients for all channels, e.g., $\vecY(\freqbinidx,\timeframeidx) = [Y_1(\freqbinidx, \timeframeidx), ..., Y_\numMics(\freqbinidx, \timeframeidx)]^T\in\mathbb{C}^\numMics$ and drop the time-frequency indices $(\freqbinidx,\timeframeidx)$ to denote the tensor with shape $(\numMics\times\numFreqs\times\numTimes)$ comprising the time-frequency points for all $\numMics$ microphones and with $\numFreqs$ and $\numTimes$ being the number of frequency-bins and time-indices respectively.

\subsection{Traditional spatial filtering}
Most traditional multi-channel speech enhancement schemes involve a spatial filter that is usually implemented following a filter-and-sum beamforming approach \cite[Sec. 12.4.2]{vary2006digital}. Such a filter-and-sum beamformer aims to suppress signal components not originating from the target direction by filtering the individual microphone signals and adding them. Using vector notation, the processing model of a filter-and-sum beamformer in the frequency domain can be formulated as 
\begin{equation}\label{eq:filterandsum}
\hat{S}(\freqbinidx, \timeframeidx) = \vech(\freqbinidx, \timeframeidx)^H\vecY(\freqbinidx,\timeframeidx) 
\end{equation}
with $\hat{S}(\freqbinidx, \timeframeidx)\in\mathbb{C}$ being an estimate of the target signal, a filter $\vech(\freqbinidx, \timeframeidx)\in\mathbb{C}^\numMics$ that may or may not be depending on the time index $\timeframeidx$ (time-variant vs. time-invariant filter) and $(\cdot)^H$ denoting the Hermitian transpose.

The simplest form is a delay-and-sum beamformer \cite[Sec. 12.4.1]{vary2006digital} that applies a filter to compensate for different time delays at the microphones caused by the differing lengths of propagation paths for the signal to reach each microphone. This approach implicitly assumes the noise signals recorded at the different microphones to be uncorrelated \cite[Sec.12.6.1]{vary2006digital}, which is a reasonable assumption for sensor noise, but not for environmental noise or interfering point sources.  

Another commonly used spatial filter is the \ac{MVDR} beamformer \cite[Sec. 12.6.1]{vary2006digital} that takes into account the correlation between microphone channels. The filter weights $\vech_{\text{MVDR}}(\freqbinidx, \timeframeidx)$ are obtained by solving the optimization problem 
\begin{IEEEeqnarray*}{rCl}\label{eq:mvdroptproblem} \vech_{\text{MVDR}}(\freqbinidx,\timeframeidx)&=&\argmin_{\vech\in\mathbb{C}^\numMics}  \vech^H(\freqbinidx,\timeframeidx)\vecPhi_V(\freqbinidx,\timeframeidx)\vech(\freqbinidx,\timeframeidx)\\ 
&&\text{s.t.}\quad \vech(\freqbinidx,\timeframeidx)^H\vecd(\freqbinidx,\timeframeidx) = 1, \qquad\IEEEyesnumber
\end{IEEEeqnarray*}
with the so-called steering vector $\vecd(\freqbinidx,\timeframeidx)$ modelling the direct path of the target signal $S(\freqbinidx,\timeframeidx)$ to the microphones and noise correlation matrix $\vecPhi_V(k,i) = \mathbb{E}[\vecV(k,i)\vecV(k,i)^H]$ with $\mathbb{E}$ denoting the statistical expectation operator. Thus, the MVDR beamformer tries to minimize the noise variance at the output of the beamformer while leaving the target signal unchanged. The latter condition is referred to as the distortionless constraint of the MVDR. The solution of the optimization problem posed in (\ref{eq:mvdroptproblem}) is given by \cite[Sec. 12.6.1]{vary2006digital}
\begin{equation}\label{eq:mvdrweights}
    \vech_{\text{MVDR}}(\freqbinidx, \timeframeidx) = \dfrac{\vecPhi_V^{-1}(\freqbinidx,\timeframeidx)\vecd(\freqbinidx,\timeframeidx)}{\vecd^H(\freqbinidx,\timeframeidx)\vecPhi_V^{-1}(\freqbinidx,\timeframeidx)\vecd(\freqbinidx,\timeframeidx)}.
\end{equation}
Adhering to the filter-and-sum processing model, and using filter weights that do not depend on the value of the noisy signal $\vecY(\freqbinidx, \timeframeidx)$ itself as in (\ref{eq:mvdrweights}), traditional spatial filtering clearly is a linear operation with respect to the noisy input. 

It has been shown that the MVDR beamformer is the optimal spatial filter under a Gaussian noise assumption \cite{balan2002MicrophoneArraySpeech, tesch2021nonlinearspatialfilteringtasl}. That is, any filter jointly performing spatial filtering and postfiltering can (in theory) be decomposed into an MVDR beamformer for spatial processing followed by a single-channel postfilter. A prominent example is the multi-channel Wiener filter, which can be decomposed in an MVDR plus single-channel Wiener filter \cite{Simmer2001}. The work by Hendriks et al. \cite{Hendriks2009multichannelMSE} and our prior work \cite{tesch2021nonlinearspatialfilteringtasl} reveal that this is not the case for more general noise distributions. The analytic filter derived in \cite{Hendriks2009multichannelMSE, tesch2021nonlinearspatialfilteringtasl} joins the spatial and spectral filtering into a non-separable non-linear operation which is in contrast to the simple and linear processing model of a beamformer. Our own previous work \cite{tesch2021nonlinearspatialfilteringtasl} demonstrates that such a joint spatial-spectral nonlinear processing may overcome the limitations of a linear beamformer, which is restricted to suppressing $M-1$ directional interfering point sources (maximum number of sources in a reverberation-free setting). However, oracle knowledge of the target and noise signals are required for accurate parameter estimation to obtain good results with the analytic joint spatial-spectral nonlinear filter. 

\subsection{DNN-based spatial filtering}
While state-of-the-art single-channel speech enhancement nowadays completely relies on DNN-based approaches, DNN-based multi-channel approaches have become a vivid research topic recently. An important step towards using the capabilities of neural networks for multi-channel speech enhancement was taken by Heymann et al. \cite{heymann2015blstm}, who design a DNN-based parameter estimation scheme for computing estimates of the steering vector and noise correlation matrix to be used in a traditional MVDR beamformer. This method has gained a reputation for its ease of use as well as good and robust results. Similarly, Togami \cite{togami2019itakura} proposes to extract speaker masks for facilitating covariance matrix and speech power estimation to be used in a multi-channel speech separation scheme. Liu et al. \cite{2018liuNeuralNetworkBased} extend the masking-based beamforming approach of \cite{heymann2015blstm} by processing multi-channel instead of a collection of single-channel inputs and providing cross-channel features. Xiao et al. \cite{xiao2016deepbeamforming} train a network to directly estimate the time-invariant filter weights $\vech(\freqbinidx)$ of a filter-and-sum beamformer from cross correlation features. %
The main drawback of a method that uses the impressive modeling capabilities of neural networks only for parameter estimation to be used in classical linear processing scheme is that the limitations of the linear model itself cannot be overcome. 

In another line of research, spatial features are used as additional input to a neural network to increase speech separation or enhancement performance, e.g.,  \cite{araki2015mcfeatures, wang2018mcdeepclustering, gu2019neuralspatialfilter, wang2019combingingspectralandspatial}. The most common spatial features are inter-channel time or phase differences (ITD/IPD), inter-channel level differences (ILD), cross-correlation based features, as well as features computed with fixed beamformers. Most of these works, show notable performance improvements over single-channel approaches proving that spatial information is very valuable for speech separation and enhancement tasks. As for all approaches using hand-crafted features, a major concern is the question whether the chosen feature design is optimal for the task at hand. For example, in \cite{zhang2021adlmvdr} and \cite{xu2021GRNNBF} the authors propose to estimate beamforming weights from speech and noise second-order statistics (covariance matrix estimates) using a \ac{DNN}, while our analysis in \cite{tesch2021nonlinearspatialfilteringtasl} suggests that higher-order statistics are a valuable source of information, which can not be exploited this way.

An increasing number of recent works skips the spatial feature design part and trains a DNN-based filter to perform speech enhancement or separation based on raw multi-channel signals, either providing the time-domain signals \cite{2020liuMultichannelSpeechEnhancement,  luo2020fasTacnet, 2019tawaraMultichannelSpeechEnhancement, 2022pandeyTPARNTriplepathAttentive, lee2021interchannel} or frequency-domain signals \cite{li2019narrowband, chakrabarty2019, tolooshams2020cadunet, wang2020complexspectralmapping, li2022eabnet, halimeh2022cospa} as input to the network. In many of these works, the authors claim that the network architecture has been designed with the goal in mind to implicitly learn a spatially selective filter from data. Nonetheless, the architectures proposed in these papers differ notably from each other. While some authors propose to learn a mask that is applied to a reference channel of the noisy signal, e.g.,  \cite{tolooshams2020cadunet, li2019narrowband}, others propose a network that outputs the real and imaginary part of the target clean speech signal \cite{wang2020complexspectralmapping} or to learn a set of coefficients $\vech$ and apply them to the signal adhering to the filter-and-sum processing model (cf. (\ref{eq:filterandsum})) \cite{luo2020fasTacnet, li2022eabnet}. %

For the last mentioned approach, it is clear that the authors have derived their architecture design from traditional linear filter-and-sum beamforming, but also others claim their architecture to be inspired be the traditional spatial filters, e.g., \cite{tolooshams2020cadunet}. The authors of EaBNet \cite{li2022eabnet} even propose to append the DNN-based spatial filter with a (DNN-based) post-filter following the traditional two-step procedure. However, it is important to be aware that their \say{spatial filter} as well as all other DNN-based approaches referenced in the last paragraph are in principle not only capable of performing non-linear spatial filtering but will likely perform spatial filtering jointly with tempo-spectral postfiltering. As a consequence, a direct comparison with a traditional linear beamformer, for example the MVDR beamformer, without a post-filter can therefore not be considered a fair comparison. 

Overall, we conclude that many interesting architectures for implementing a DNN-based filter for multi-channel speech enhancement have been proposed, but also a lot of open questions remain. Most approaches haven been evaluated with respect to their overall speech enhancement performance. However, this is not very informative with regard to the internal mechanisms of the network. For example, it is unclear whether a network architecture inspired from traditional beamforming performs particularly well in spatial filtering as hypothesized by many authors, since performance improvements could also be achieved by better exploitation of tempo-spectral information. Thus, a more systematic evaluation is required to provide insights in the internal mechanisms of these DNN-based filters. 

\section{Proposed approach}\label{sec:netarcs}
In this work, we aim to investigate the contribution of different sources of information, that is spatial, spectral and temporal information, to a \ac{DNN}-based filter for a speech enhancement or speech extraction problem. We are particularly interested in understanding the nature of a non-linear spatial filter and its interdependencies with temporal and spectral information. To provide insights into the \say{black box} of a DNN-powered filter, we use a simple network structure that allows us to easily control the integration of different sources of information and a dataset that makes it easy to assess the quality and properties of the spatial filter. This section describes the network design used in our experiments.

\subsection{Base network architecture (F-JNF, T-JNF)}\label{subsec:basenetwork}
\begin{figure*}[htb]
    \centering
    \includegraphics{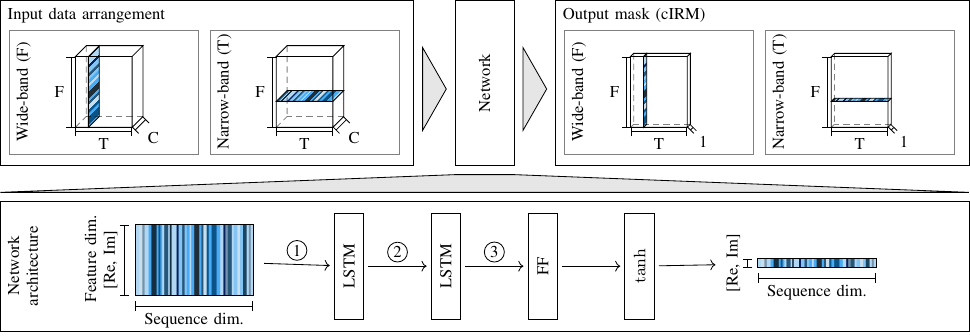}
    \caption{Illustration of the base system architecture. The input data is arranged according to a wide-band or narrow-band input and fed into a network with two LSTM layers, an FF layer and $\tanh$ activation to obtain an estimate of a cIRM. }
    \label{fig:base-arc}
\end{figure*}
For our experiments, we adapt the architecture proposed by Li and Horaud \cite{li2019narrowband, 2020liNarrowbandDeepFiltering}, that performs speech enhancement using a mask estimated from narrow-band multi-channel inputs. The distinctive feature of their approach is that the network processes all frequency bands separately. The network weights, however, are shared between frequencies. In the following, we propose a number of alternative network architectures to enable a detailed analysis. Figure \ref{fig:base-arc} depicts the base network architecture. As can be seen in the bottom part, the network consists of only three layers, two (bi-directional) \ac{LSTM} layers followed by a \ac{FF} layer. An LSTM layer \cite{hochreiter1997lstm} is commonly used for sequence modeling. In our setup, the feature dimension (vertical) mostly corresponds to channel information (real and imaginary parts stacked) while the sequence dimension (horizontal) is chosen according to the second source of information which could be time (narrow-band) or spectral (wide-band) information. As spatial information is processed jointly with a second source of information, we denote this network as joint non-linear filter (JNF) prefixed with T or F in the narrow-band and wide-band case respectively. Thus, the narrow-band version (T-JNF) as proposed in \cite{li2019narrowband} has access to fine-grain spatial and temporal information but only global spectral statistics, while our proposed variant F-JNF can leverage fine-grained spectral information in addition to spatial information.

\subsection{Combining temporal and spectral information (FT-JNF)}\label{subsec:FTJNF}
The basic architecture described in Section \ref{subsec:basenetwork} combines spatial information with spectral \emph{or} temporal information. Next, we propose a variant that can exploit all three sources of information combining spatial with tempo-spectral processing. In order to ensure comparability of the results, we do not change the basic architecture or the number of parameters. Instead, we manipulate the data arrangement at the position marked with a circled two. The filter denoted by FT-JNF then feeds wide-band data into the first LSTM layer. The obtained features are then switched to a narrow-band arrangement before input to the second LSTM layer. This way, the FT-JNF can potentially exploit all three sources of information. %

\subsection{Non-linear spatial filtering (T-NSF, F-NSF, FT-NSF)}\label{subsec:NSF}
To study the properties of a non-linear spatial filter (NSF) separately from the tempo-spectral processing, we define three additional variants of the the network architecture: T-NSF, F-NSF, and FT-NSF. The underlying idea is to prevent the network from employing fine-grained temporal and spectral information by randomly permuting the data along the sequence dimension before feeding it into the LSTM at position \ding{192}. The inverse permutation operation is then applied before the FF layer at position \ding{194}. Accordingly, only global statistics with respect to the frequency or time dimension are available but correlations between neighboring frequencies or time steps cannot be exploited. Preliminary experiments have shown that the spatial processing using a wide-band data arrangement (F-NSF) performs poorly if the frequency-bin index is unknown to the network. This is likely because the spatial characteristics of the data depend strongly on the frequency-bin index. To ensure that this information is still available after shuffling along the sequence dimension, we append the frequency-bin index to the feature dimension. We do this also for a narrow-band data arrangement (T-JNF) but in this case the effect on the performance is minor. 
Analogous to the procedure described in Section \ref{subsec:FTJNF}, we define a non-linear spatial filter that incorporates both global spectral and global temporal information. This is achieved by again switching from wide-band to narrow-band data arrangement at position \ding{193} and requires both LSTM layers to be wrapped in permutation and inverse permutation operations with respect to the respective sequence dimension.

\subsection{DNN-based post-filtering (PF)}\label{subsec:post-filter}
Finally, we introduce a single-channel post-filter that jointly processes temporal and spectral information. For consistency, we stick to the simple base architecture shown in Figure \ref{fig:base-arc}. Here, the real and imaginary parts of the single-channel input data are stacked along the frequency dimension to form the feature dimension. The time axis is then used as sequence dimension.

\subsection{Lossfunction and training details}\label{subsec:trainingdetails}
We train the networks based on a \ac{cIRM} \cite{williamson2016cirm} in favor of a magnitude \ac{IRM} as in \cite{li2019narrowband} to facilitate phase enhancement. For this reason, the \ac{FF} layer is followed by a \emph{tanh} activation function, which outputs a compressed mask estimate. We use compression parameters $K=C=1$ as defined in \cite{williamson2016cirm}. The enhanced signal is then obtained by multiplication of the uncompressed target speech \ac{cIRM} $\mathcal{M}_\text{S}(k,i)\in\mathbb{C}$ with the noisy recording $Y_0(k,i)$ using the first channel as reference, i.e., 
\begin{equation}\label{eq:speechest}
    \hat{S}(k,i) = \mathcal{M}_\text{S}(k,i) \cdot Y_0(k,i).
\end{equation}
The real and imaginary parts of the noise \ac{cIRM} $\mathcal{M}_{\text{V}}$ can be obtained from the real and imaginary part of the target speech \ac{cIRM} using \cite{tolooshams2020cadunet}: 
\begin{IEEEeqnarray}{rCl} 
\Re(\mathcal{M}_\text{V}) &=& 1 - \Re(\mathcal{M}_\text{S}),\\ 
\Im(\mathcal{M}_\text{V})&=&-\Im(\mathcal{M}_\text{S}).
\end{IEEEeqnarray}
The noise \ac{cIRM} estimate can be used to obtain an estimate of the pure noise component contained in the signal, i.e, 
\begin{equation}\label{eq:noiseest}
    \hat{V}(k,i) = \mathcal{M}_\text{V}(k,i) \cdot Y_0(k,i).
\end{equation}

We use the loss function proposed by Tolooshams et al. \cite{tolooshams2020cadunet}, which is composed of time and frequency domain $\ell_1$ loss terms: 
\begin{equation}\label{eq:loss}
    L(s, \hat{s}) = \sum_{u\in\{s, v\}} \alpha\norm{u-\hat{u}}_1 + \norm{|U|-|\hat{U}|}_1.
\end{equation}
Here, the frequency-domain terms $\hat{S}$ and $\hat{V}$ are estimated as given in (\ref{eq:speechest}) and (\ref{eq:noiseest}) and time-domain quantities are obtained by an inverse \ac{STFT}. We set $\alpha=10$ to equalize the contribution of either domain in the loss term. 

As can be seen from the loss function, our training scheme uses the noisy observations $\vec{y}(t)$, which serves as network input, as well as the ground truth noise signals $\vec{v}(t)$ recorded at the microphones and the non-reverberant signal $s(t)$, which has been aligned with the noisy observation to include the propagation delay. If the ground truth for the noise signal is unknown, we only use the clean speech related parts of the loss function. During training, we randomly extract three seconds of audio from an utterance and compute the \ac{STFT} using a 32 ms long window with 50\% overlap. The $\sqrt{\text{Hann}}$ window is applied for analysis and synthesis. We train the networks with batch size six until convergence with maximum 250 epochs and select the best model with respect to the validation loss. The number of LSTM units is set to 256 and 128 for all networks, except PF, for which 256 units are used in both layers. The Adam optimizer \cite{KingmaB2015Adam} with learning rate 0.001 is used. 

\section{Analysis of the interplay of spatial with tempo-spectral information}\label{sec:mixeval}
In this section, we evaluate the previously described networks in a speaker extraction scenario with a single speaker that is to be extracted and five additional interfering speech sources. Such a scenario seems particularly suitable to study the spatial filtering capabilities of a processing method since a spatially selective filter, as opposed to a filter that mainly exploits tempo-spectral information, is expected to be the key to good performance on this task. This is because the target signal has very similar tempo-spectral properties as the interfering signal (five speakers) but the signals differ decisively in their spatial properties.

\subsection{Dataset}
\begin{table}
    \caption{For each sample, the room characteristics are obtained by sampling uniformly from the value ranges given in this table.}
    \label{table:simu}
    \centering
    \begin{tabular}{cccc}\toprule
         Width & Length & Height & T60 \\\midrule
          $2.5-5$ m &  $3-9$ m  &  $2.2-3.5$ m   & $0.2 - 0.5$ s \\\bottomrule
    \end{tabular}
\end{table}
\begin{figure}[tb]
\centering
\includegraphics{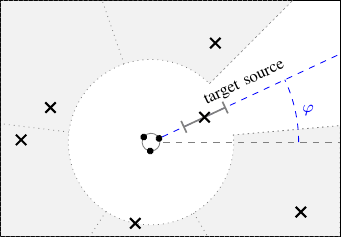}
    \caption{Illustration of the simulation setup. The target source is located in a fixed orientation with respect to microphone array. The five interfering sources are placed in the gray area (one per segment). Room properties are sampled from the given ranges.}
    \label{fig:simu}
\end{figure}
We generate a simulated dataset using \texttt{pyroomacoustics} \cite{scheibler2018pyroomacoustics}, which provides an implementation of the source-image model \cite{allen1979image}. The setup is illustrated in Figure \ref{fig:simu}. For each sample, the room dimensions and the reverberation time are uniformly sampled from the value ranges given in Table \ref{table:simu}. We use a circular microphone array with a diameter of $10$ cm and between two and five channels. The microphone array is placed at a random position in the xy-plane but at least $1$ m away from the walls, and it is located at a height of $1.5$ m. As depicted in Figure \ref{fig:simu}, a rotation $\varphi$ is applied to the microphone array sampled from the interval $[0,2\pi)$. 
In our setup, the target speaker has to be identified by its spatial location. Accordingly, we place the target speaker in a fixed position relative to the microphone orientation on the blue dotted line in Figure \ref{fig:simu}. Its distance to the microphone array ranges between $0.3$ m and $1$ m. The five interfering sources are placed in the gray area with a minimum distance of $1$ m to the microphone array location. As indicated by the white area, a room spanned by the $20$\degree~angle to either side of the target source is also kept free of interferers. To ensure an even distribution of sources in the room, we place one interfering source per segment as indicated by the dashed gray lines. The height of the interfering speech sources is sampled from a normal distribution with mean $1.6$~m and standard deviation $0.08$.

We generate 6000, 1000, and 600 samples with a sampling frequency of $16$~kHz for training, validation and testing respectively using clean speech signals from the WSJ0 dataset \cite{wsjdata2007}. Signals between the different sets do not overlap. The \ac{SNR} is not explicitly controlled but obtained from the the simulation setup with varying distances of the sources to the microphone array. The average SNR is $-4$~dB and $95$\% of the data samples distribute between $-9$~dB and $2$~dB SNR.

\subsection{Separability of spatial processing and post-filtering}\label{subsec:evalseparability}
Figure \ref{fig:1-separated} illustrates the traditional two-step approach with a spatial filter that is applied first and a single-channel post-filter for tempo-spectral processing that is applied in a second processing step. Such a modular design is desirable as it offers flexibility and interpretability, however, the analytical \ac{MMSE} solution in a non-Gaussian noise scenario is non-linear and non-separable \cite{tesch2021nonlinearspatialfilteringtasl}. The MMSE-optimal solution thus corresponds to the joint spatial and (tempo-)spectral filter depicted in Figure \ref{fig:1-jointfilter}. However, it is unclear if the third option of using a non-linear spatial filter as depicted in Figure \ref{fig:1-separated2} is a meaningful concept or if non-linear spatial processing is only useful if tempo-spectral information and spatial information are processed jointly as in Figure~\ref{fig:1-jointfilter}. For this reason, in this section, we investigate if a DNN-based non-linear filter can be separated into spatial processing and single-channel tempo-spectral post-filtering by comparing the performance of all three configurations shown in Figure \ref{fig:1-comparison}. 
\begin{figure}
  \centering
    \includegraphics{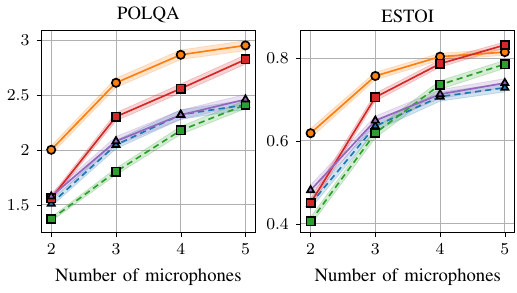}
    \includegraphics{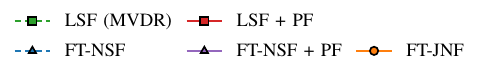}
  \caption{We report the mean POLQA and ESTOI scores along with the 95\% confidence interval for a set of multi-channel filters. This figure shows that joint spatial and tempo-spectral filtering (FT-JNF) outperforms a nonlinear spatial filter plus a postfilter (FT-NSF+PF).}\label{fig:2-nummics}
\end{figure}
\begin{figure}[t]
    \centering
    \includegraphics{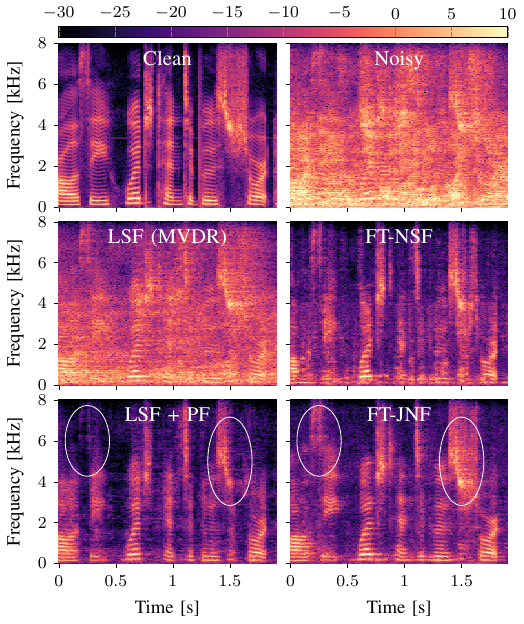}
    \caption{Spectrogram visualization of an example utterance. The target signal and the noisy observation are displayed in the top row. The middle row shows two spatial filters, a linear MVDR on the left and a DNN-based non-linear on the right. The bottom depicts the MVDR with an independent post-filter and the joint spatial and tempo-spectral filter.
    }
    \label{fig:spectrograms}
\end{figure}

The left plot in Figure \ref{fig:2-nummics} shows the mean \ac{POLQA} score \cite{polqa2018} and the $95$\% confidence interval for a varying number of microphones. The \ac{POLQA} algorithm is the successor to the \ac{PESQ} measure \cite{pesq2007}. It measures speech quality based on \ac{MOS} scale ranging from one (bad) to five (excellent). The dashed lines correspond to spatial-only filters. That is the traditional \ac{MVDR} beamformer (green) and the FT-NSF described in Section \ref{subsec:NSF} (blue). The parameters of the \ac{MVDR} beamformer are estimated from oracle data. We compute the time-varying noise covariance estimate by recursive averaging of the pure noise data and estimate the \ac{ATF} by multiplying the principal eigenvector of the generalized eigenvalue problem for speech and noise covariance matrices with the speech covariance matrix as described in \cite{ito2017}. 

Even though the \ac{MVDR} parameters were accurately estimated from oracle data, which means that the \ac{MVDR} should be considered as an upper bound on the spatial filtering performance achievable with a linear processing model, the non-linear spatial filter excluding a tempo-spectral post-filtering yields higher \ac{POLQA} scores, in particular for a small number of microphones. A spectrogram visualization for three microphones is shown in Figure \ref{fig:spectrograms}. The results obtained with a linear spatial filter (LSF) and a non-linear spatial filter (FT-NSF) are depicted in the middle row. Differences in the behavior are clearly visible: While the \ac{MVDR} is distortionless by design at the cost of little noise suppression in this difficult noise scenario, the non-linear spatial filter aggressively reduces noise, but introduces quite some speech distortions. Please find audio examples on our website\footnote{\url{https://uhh.de/inf-sp-deep-non-linear-filter}}. 
Next, we combine each spatial filter with an independent single-channel post-filter. For this, the \ac{DNN} described in Section \ref{subsec:post-filter} is trained using the output of the \ac{MVDR} and FT-NSF evaluated on the training set as network input. The results for these two-step approaches are shown in Figure \ref{fig:2-nummics} using the same marker as the corresponding spatial filter but with a solid line. We find that the post-filter added to the non-linear spatial filter (FT-NSF+PF) does not result in a notable performance improvement. In effect, the purple line runs almost exactly on top of the blue dashed line. This can be explained by the fact that speech information, which was lost already during spatial processing, cannot be recovered by multiplication with the post-filter mask. In contrast, the \ac{MVDR} beamformer does not distort the clean speech signal, and adding a single-channel post-filter, represented by the red solid line, is very effective. Here, we observe a performance boost between 0.18 POLQA score (two microphones) and 0.5 \ac{POLQA} score (three microphones) in comparison with the linear spatial filter only. %

Finally, we compare with the joint non-linear spatial and tempo-spectral filter FT-JNF. As visible in Figure \ref{fig:1-jointfilter}, the separation of spatial and tempo-spectral processing has been removed, which allows the network described in Section \ref{subsec:FTJNF} to exploit the interdependencies between spatial and tempo-spectral information. This joint approach, depicted by the solid orange line, clearly outperforms the separated linear spatial filter plus post-filter approach for a low number of microphones. For two microphones the difference even amounts to 0.44 \ac{POLQA} score. With an increased number of microphones, the gap between the orange line (FT-JNF) and the red line (LSF+PF) decreases or inverts even for ESTOI \cite{2016jensen} depicted in the right plot. This is not very surprising as the number of anechoic point sources that can be canceled by the oracle \ac{MVDR} increases by one for every added microphone. Accordingly, the performance of the spatial filter improves considerably with every microphone added, and when combined it with a strong post-filter, it becomes increasingly difficult to outperform the \emph{oracle} \ac{MVDR} plus post-filter with a data-driven filter. 

Overall, two conclusions emerge from these results: First, the joint non-linear spatial and tempo-spectral filter (orange) drastically outperforms the non-linear spatial filter with an independent post-filter (purple) in terms of speech quality and intelligibility. This means that the dependencies between spatial and tempo-spectral information are successfully exploited by the neural network.
And second, the DNN-based joint non-linear filter (FT-JNF) significantly outperforms the oracle \ac{MVDR} with an added single-channel post-filter for a small number of microphones. 

\subsection{Interdependency of spatial processing with spectral and temporal information}\label{subsec:evalinterdependency}

The experiment in the previous section demonstrated that spatial processing should not be separated from tempo-spectral processing, as these two seem to mutually enrich each other. In this section, we will further investigate the interdependencies between spatial processing and temporal and spectral processing. 

\newcolumntype{m}[1]{D{:}{\pm}{#1}}
\begin{table}
    \caption{Impact of different sources of information (spectral (F) and temporal (T)) used besides spatial information. We report mean improvements and the 95\% confidence interval.}
    \label{table:contribution}
    \centering
    \begin{tabular}{l@{}cm{-1}@{}m{-1}}\toprule
         &~~&  \multicolumn{1}{c}{$\Delta$ POLQA} & \multicolumn{1}{c}{ESTOI }\\\midrule
          F-NSF && 0.78~:~0.03 & 0.62~:~0.012\\
          T-NSF && 0.46~:~0.03 &  0.54~:~0.013\\
          FT-NSF && 0.87~:~0.03 & 0.64~:~0.011\\\midrule
          F-JNF && 1.15~:~0.04 & 0.70~:~0.011\\
          T-JNF \cite{li2019narrowband} && 0.74~:~0.03 & 0.63~:~0.012\\
          FT-JNF (proposed) && \textbf{1.43}~:~\textbf{0.04} & \textbf{0.76}~:~\textbf{0.009}\\\bottomrule
    \end{tabular}
\end{table}

In the top three rows of Table \ref{table:contribution}, we report the results obtained with a non-linear spatial filter that has access to global spectral, temporal or tempo-spectral information using three microphones. The corresponding neural network architectures have been explained in Section \ref{subsec:NSF}. As expected, we observe that the highest performance is obtained with a non-linear spatial filter that incorporates both, global temporal and spectral information, denoted by FT-NSF. However, the comparison of F-NSF and T-NSF reveals that spatial processing here benefits much more from global spectral than global temporal information. The difference even amounts to 0.32 POLQA score and is also reflected in the ESTOI measurements. A similar pattern is also observed for the joint non-linear filter that can not only exploit global statistics but also fine-grained information including correlations between neighboring frequency bins and/or time steps. The performance differences between F-JNF and T-JNF amount to 0.41 POLQA score and $0.07$ ESTOI score. 

\begin{figure}
  \centering
    \includegraphics{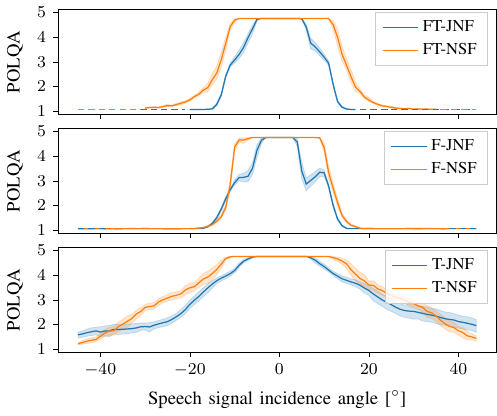}
  \caption{Visualization of the spatial selectivity of the learned filters. The plots show the the mean POLQA score and 95\% confidence interval for a clean and anechoic signal arriving from a given incidence angle. A low POLQA score here corresponds high suppression of the signal, while a very high POLQA score (around $0\degree$) means that the signal has passed through the filter unaltered. Signals for which no POLQA score can be computed are marked with a dashed line.}\label{fig:speechangle}
\end{figure}

The impact of different sources of information on the spatial selectivity of the filter is visualized in Figure \ref{fig:speechangle} in more detail. For this, we present the trained networks with a clean speech signal originating from varying directions with $1$ m distance from the microphone array. For this experiment, we use a simulated anechoic room as we want to measure the filter's response to a signal from a specific direction. The plots in Figure \ref{fig:speechangle} show the \ac{POLQA} score for the filtered signals averaged over 15 examples. A high POLQA score, which is attained by all filters near $0$\degree, corresponds to a signal that has passed through the filter unaltered, while a low POLQA score indicates high suppression of the signal. The POLQA algorithm does not provide a result if the signal is not speech-like anymore and has very low energy. For these processed signals, which retain less then $0.1$\% of their original energy, we indicate high suppression with a dashed line at the minimum POLQA score.

Comparing the two bottom plots of Figure \ref{fig:speechangle}, it is clearly visible that exploiting frequency information as opposed to time information increases the spatial selectivity, which can serve as an explanation of the performance differences observed before. While all plots show a \say{distortionless} response for signals with an incidence angle between $-4$\degree~and $4$\degree, signals arriving from a larger angles are much less suppressed (resulting in a higher \ac{POLQA} score) for the network using temporal information. In particular, even signals that arrive from the interference region are not fully suppressed.  Furthermore, considering the upper two plots, it is interesting to observe that adding fine-grained spectral information in FT-JNF and F-JNF narrows down the spatial selectivity even beyond the $-20$\degree~and $20$\degree~angle that can be expected from the dataset configuration. Yet, a narrower selectivity pattern might be helpful to resolve the spatial characteristics in a noisy scenario.

\section{Comparison to state-of-the-art methods}\label{sec:sotaeval}
In Table \ref{table:contribution}, the overall best performance is obtained with the joint non-linear filter FT-JNF that exploits tempo-spectral in addition to spatial information. Comparing to T-JNF which has originally been proposed by Li and Horaud \cite{li2019narrowband}, we find that our systematic evaluation of the interplay between spatial and temporal as well as spectral information leads to a drastic performance improvement of 0.69 POLQA score in a speaker extraction scenario.

\subsection{Baselines}\label{subsec:baselines}
In this section, we compare the proposed FT-JNF with four additional baseline network architectures besides T-JNF. This ensures that the study we conducted with a rather simple network provides meaningful results also in comparison with recent and more elaborate state-of-the-art network architectures and it furthermore allows us to assess the question whether a network design inspired by a traditional filter-and-sum beamformer, e.g., \cite{luo2020fasTacnet, li2022eabnet, halimeh2022cospa}, is likely to exhibit enhanced spatial filtering capabilities. %

As our primary focus in this work is to better understand the consequences of architectural choices for implementing multi-channel DNN-based filters, we train all baseline architectures following the same procedure outlined in Section \ref{subsec:trainingdetails} and using the loss function defined in (\ref{eq:loss}). For most baselines, we use the code provided by the authors with the default parameter settings and focus our parameter search mostly on the learning rate. The selected values are given in Table \ref{table:mix-sota-config}. It is likely that an extensive hyper-parameter tuning might lead to better results, but we nevertheless consider the results representative of their respective network architecture on the used dataset. Deviations from the training procedure or the settings described in the respective paper will be noted in the following. These are the baselines that we compare the proposed FT-JNF to: 

\begin{itemize}
\item T-JNF: We consider the architecture T-JNF as an instance of the network proposed by Li and Horaud \cite{li2019narrowband}. However, in order to facilitate phase processing, we have changed the network output from \ac{IRM} to \ac{cIRM} and also replaced the final output layer with a tanh layer accordingly. 
\item CRNN: We reimplement a variant of the \ac{CRNN} for mask estimation proposed by Chakrabarty and Habets \cite{chakrabarty2019}. The authors propose a \ac{CNN} for spatial feature extraction. For this small convolution kernels are used on the channel dimension such that a series of convolutional layers reduces the channel dimension to one. These spatial features are then processed with a bi-directional \ac{LSTM} and fed into a \ac{FF} layer to produce a mask. We use real and imaginary parts as input and estimate a \ac{cIRM}.
\item FaSNet+TAC: FasNet \cite{luo2019fasnet} is a time-domain approach mimicking a traditional filter-and-sum beamformer. %
The authors proposed an extension, denoted FaSNet+TAC \cite{luo2020fasTacnet}, which enables variable microphone array configurations. As the authors report improved performance also for fixed array geometries, we choose to evaluate FaSNet+TAC on the speaker extraction dataset. We use the implementation provided by the authors. 
\item EaBNet: Li et al. \cite{li2022eabnet} propose the Embedding and Beamforming Network (EaBNet). It uses a U-Net structure to estimate an embedding that incorporates spatial and tempo-spectral information and then employs a \say{beamformer} network to obtain weights that are applied in a filter-and-sum beamforming manner. We use the implementation provided by the authors using the LSTM branch. We do not apply a single-channel \ac{DNN} (post-filter network) to the output of EaBNet and use uncompressed network inputs and targets. This baseline uses shorter STFT windows of length $20$ ms and 50\% overlap. 
\item COSPA: The Complex-valued Spatial Autoencoder (COSPA) has been proposed by Halimeh and Kellermann \cite{halimeh2022cospa}. Similar to EaBNet it adopts a filter-and-sum approach with frequency-domain complex-valued coefficients estimated by the network. The network architecture is composed of an encoder, a compandor and a decoder part. All of these are complex-valued networks. %
We use the implementation provided by the authors, which uses $64$ ms long STFT windows and an overlap of $50$\%. We train using the clean speech terms in the loss function only.
\end{itemize}

\subsection{Performance analysis}

\begin{figure}[t]
    \centering
    \includegraphics{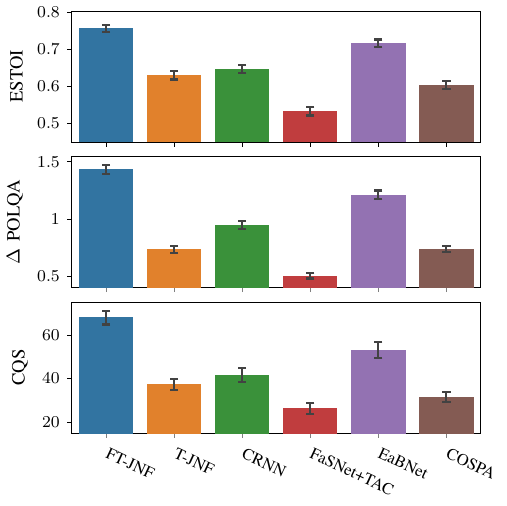}
    \caption{Performance comparison of the proposed architecture FT-JNF and five baselines. The two upper plots show the mean ESTOI and POLQA performance on the speaker extraction dataset and the 95\% confidence interval. The bottom plot shows the CQS results obtained by a MUSHRA listening experiment on twelve randomly selected examples.}
    \label{fig:mix-sota}
\end{figure}
\begin{table}
    \caption{Baseline configurations.}
    \label{table:mix-sota-config}
    \centering
    \begin{tabular}{l@{\hskip.3cm}l@{\hskip.3cm}l@{\hskip.2cm}l@{\hskip.2cm}l}\toprule
         &LR& STFT & \#Param. & Implementation /\\
         &&[ms]&  [M]& Github repository\\\midrule
          FT-JNF & 0.001& 32 & 1.2 & own \href{https://github.com/sp-uhh/deep-non-linear-filter}{(sp-uhh/deep-non-linear-filter)}\\
          T-JNF & 0.001& 32 & 1.2 & own\\
          CRNN &0.0001& 32 & 17.4 & own\\
          FaSTAC &0.0001& -- & 4.1 & \href{https://github.com/yluo42/TAC}{ylou42/TAC}\\
          EaBNet &0.001& 20 & 2.8 & \href{https://github.com/Andong-Li-speech/EaBNet}{Andong-Li-speech/EaBNet}\\
          COSPA &0.0001& 64 & 2.1 & \href{https://github.com/ModarHalimeh/COSPA}{ModarHalimeh/COSPA}\\\bottomrule
    \end{tabular}
\end{table}

We train and evaluate all networks on the speaker extraction dataset. The results with respect to the \ac{POLQA} improvements and ESTOI scores are displayed in the two upper plots of Figure \ref{fig:mix-sota}. Here, we observe that the proposed FT-JNF consistently outperforms all other baselines by at least 0.22 \ac{POLQA} score and 0.04 ESTOI score. In addition to using objective performance measures, we also conducted a MUSHRA \cite{mushra2015} listening experiment with eleven participants using the webMUSHRA framework \cite{schoeffler2018webmushra}. The participants have rated the overall quality of the algorithms based on twelve randomly sampled examples. The results are reported on a \ac{CQS} and presented in the bottom plot. The test results align very well with the objective measures and we find that FT-JNF performs best with a score of $67.9$ outperforming EaBNet in second place with a score of $53.1$. This is despite the fact that our proposed FT-JNF has the least number of learnable parameters. The number of parameters for each network architecture are given in Table \ref{table:mix-sota-config}. It is apparent that the number of parameters is not the decisive factor for good performance here. Since all networks were trained in the same way (data, loss, optimizer etc.), we attribute the performance differences to the architectural choices of how to integrate different sources of information in the processing. 

While the architectures described in Section \ref{sec:netarcs} as well as the CRNN adopt a mask-based approach, the baselines FasNet+TAC, EaBNet and COSPA resort to the filter-and-sum technique from traditional beamforming, where the filter weights are learned by the respective network. As the speaker extraction dataset is very challenging with low SNR and many interfering speech sources that have a similar tempo-spectral structure as the target signal, we can interpret the results in Figure \ref{fig:mix-sota} to reflect to a large extend the spatial selectivity of the DNN-based filters. Contrary to the common belief that a network design guided by the traditional beamforming paradigm is beneficial to spatial filtering capabilities, the best performance is obtained by FT-JNF that employs a mask-based approach, while the beamformer-inspired EaBNet only performs second best with an audible performance difference.
\begin{figure}[t]
    \centering
    \includegraphics{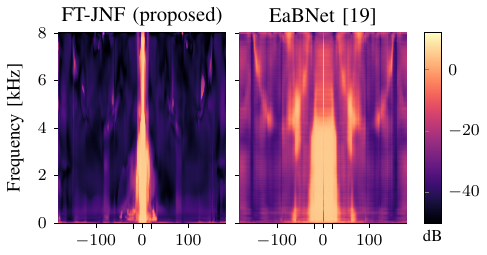}
    \caption{Visualization of the spatial selectivity of the learned filters. The patterns are created by presenting white noise signals to the networks and averaging the resulting STFT signal along the time dimension for each incidence angle and converting to decibel.}
    \label{fig:mix-sota-pattern}
\end{figure}
\begin{figure}[t]
    \centering
    \includegraphics{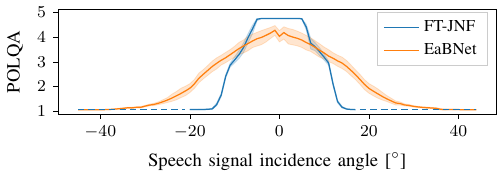}
    \caption{Visualization of the spatial selectivity of the learned filters. The plots show the the mean POLQA score and 95\% confidence interval for a clean and anechoic signal arriving from a given incidence angle.
    }
    \label{fig:mix-sota-angle-speech}
\end{figure}

In order to investigate further the spatial selectivity of the different approaches, we perform an experiment similar to the one presented in Figure \ref{fig:speechangle}. Here, we present the trained networks FT-JNF and EaBNet with spectrally white noise signals originating from variable directions in an anechoic room. Clearly, these signals are out-of-distribution data for a network trained on speech mixtures. However, the spatial properties are still consistent with the ones the network has seen during training. Figure \ref{fig:mix-sota-pattern} displays the filters' response to  these spatial cues. The incidence angle of the white noise signals is plotted on the x-axis. For each direction the \ac{STFT} of eight filtered signals are averaged along the time axis. These white-noise response patterns seem to resemble the traditional directivity patterns \cite[Sec. 12.5.2]{vary2006digital}. However, it must be noted that these white-noise response patterns do not allow for the same interpretation as a traditional beampattern. The reason for this is the non-linearity of the DNN-based filters. While a traditional beamformer, due to its linear nature, can in principle process all directional components of a signal separately and compose the final result after processing, this is not possible for a non-linear approach. 

Bearing this in mind, the plots in Figure \ref{fig:mix-sota-pattern} nevertheless provide interesting insights into the spatial processing performed by the two networks. The FT-JNF shows a very clear spatial selectivity oriented towards the known position of the target source at zero degree. The width of the beam here coincides quite well with the two additional ticks at $-20$\degree~ and $20$\degree, which mark the noise-free spatial section. On the other hand, the beam produced by EaBNet is much wider and suppression in the non-target direction does not work as well in particularly for high frequencies. What is is also noticeable is that the pattern suggest that signals near zero degree are slightly low-pass filtered, while the signal originating from an exact zero degree angle is high-pass filtered to some extend. 

This loss in overall signal quality is also visible in Figure~\ref{fig:mix-sota-angle-speech} for EaBNet. Here, we repeat the previously described experiment, where we present clean speech signals from different directions as input to the network (Figure \ref{fig:speechangle}). Comparing the orange line representing EaBNet with the blue line for FT-JNF, we find that EaBNet reduces the quality of the clean speech signal even if it is presented from the target direction. Considering this and also the width of the beam in both figures, we conclude that the performance differences that we have found in Figure \ref{fig:mix-sota} are well explained by the spatial properties of the filters. 

\section{Implications of training DNN-based multi-channel filters on CHiME3}\label{sec:chimeeval}
Finally, we evaluate on the CHiME3 data \cite{barker2015chime}, which has been recorded in four real-world noisy environments: a cafeteria, a bus, a pedestrian area and next to a busy street. This dataset is frequently used to train and evaluate DNN-based multi-channel algorithms. The recordings have been conducted with a six-channel microphone array attached to a tablet that is held by the recorded speaker. 

\subsection{Dataset}
The T-JNF network proposed by Li and Horaud \cite{li2019narrowband} has originally been trained on the CHiME3 data. The authors propose  in \cite{li2019narrowband} to create a simulated dataset, which combines the pure noise recordings provided in the CHiME3 dataset with clean booth recordings instead of artificially spatialized target signals. We use their data generation script to obtain 2400, 476, and 3251 utterances for training, validation and test respectively. The signals in the test set are mixed with a SNR in $\{-4,0,4,8\}$ dB and we use the last four channels for our experiments.

\subsection{Performance analysis}
First, we assess the interaction between spatial and spectral as well as temporal information also on the CHiME3 dataset. Therefore, in Table \ref{table:chime-results}, we report the POLQA improvement scores for FT-JNF, F-JNF and T-JNF. As before and as expected, we find that the best performance is obtained by FT-JNF in the top row that can exploit all available sources of information, that is spatial, spectral and temporal information. However, a comparison with the bottom part of Table \ref{table:contribution} showing results for the speaker extraction dataset reveals that the performance benefit of including spectral versus temporal information is reversed here. While a spatial-spectral filter performs better on the speaker extraction dataset, a spatial-temporal filter prevails on the CHiME3 dataset even though with a smaller performance gap. This behavior can be explained by considering the differences in the signal characteristics of the two datasets. While the speaker extraction dataset requires high spatial selectivity for good performance, which means that multi-channel processing is required, a single-channel filter performing tempo-spectral enhancement is expected to obtain solid results on the CHiME3 dataset. This is because the noise signals in the CHiME3 dataset have a tempo-spectral structure that is quite different from that of the target speech signal and are, in most cases, much more stationary. 

\begin{figure*}
    \centering
    \includegraphics[width=\textwidth]{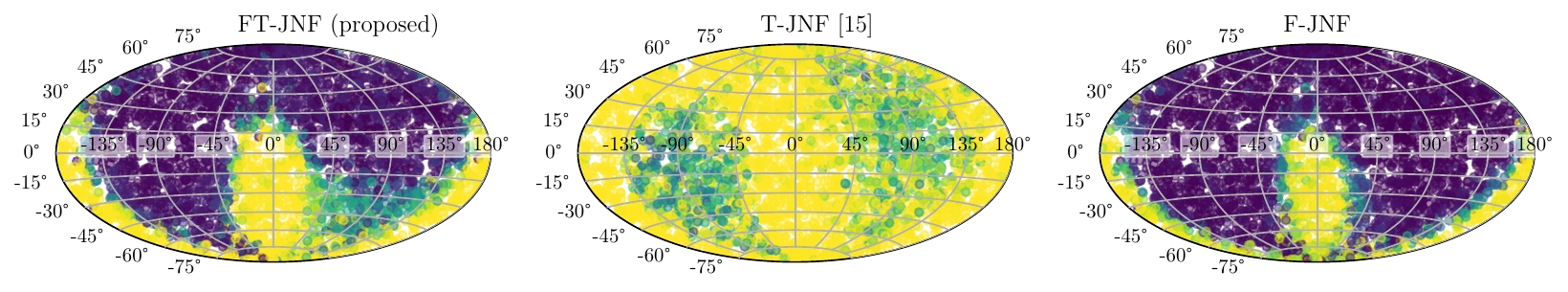}
    \includegraphics{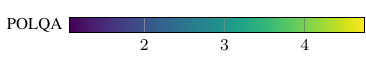}
    \caption{Spatial selectivity maps for filters trained on the CHiME3 data. The scatter plots show the POLQA scores for a clean and anechoic signal arriving from a given incidence angle. Examples for which POLQA could not be computed because there is so little energy retained in the signal were assigned the minimum POLQA score. The data points, which lie on a sphere of radius 40 cm, are projected into the plane using the Hammer projection.}
    \label{fig:chime_selectivity}
\end{figure*}
\begin{table}
    \caption{POLQA improvement scores (mean and 95\% confidence interval) for the proposed network architectures and baselines evaluated on the CHiME3 data.}
    \label{table:chime-results}
    \centering
    \begin{tabular}{l@{}m{-1}@{}m{-1}@{}m{-1}@{}m{-1}}\toprule
         & \multicolumn{1}{c}{BUS} & \multicolumn{1}{c}{CAF} & \multicolumn{1}{c}{PED} & \multicolumn{1}{c}{STR}\\\midrule
          F-JNF & 1.16:0.05 & 1.17:0.05 & 1.08:0.04 & 1.35:0.03\\
          T-JNF & 1.30:0.03 & 1.23:0.03 & 1.11:0.03 & 1.45:0.03\\
          FT-JNF & \textbf{1.53}:\textbf{0.04} & \textbf{1.56}:\textbf{0.04} & \textbf{1.45}:\textbf{0.04} & \textbf{1.76}:\textbf{0.03}\\\midrule
          CRNN & 0.89:0.04 & 0.90:0.04 & 0.83:0.04 & 1.02:0.03\\
          FaSNet+TAC & 0.61:0.03 & 0.53:0.03 & 0.51:0.02 & 0.61:0.02\\
          EaBNet & 1.19:0.04 & 1.18:0.04 & 1.08:0.04 & 1.31:0.03\\
          COSPA & 0.60:0.03 & 0.61:0.03 & 0.56:0.03 & 0.65:0.03\\\bottomrule
    \end{tabular}
\end{table}

Consistent with the results of Section \ref{subsec:evalinterdependency}, in Figure \ref{fig:chime_selectivity} we show that a spatio-temporal filter (T-JNF) has a substantially lower spatial selectivity than a spatio-spectral filter: The plots have been obtained by providing the network trained on the CHiME3 data with a clean speech input from a variable direction. For this, we simulate the CHiME3 microphone array in a room with a clean speech source in a variable position with $40$ cm distance to the microphone array. Signal suppression (blue) or signal pass-through (yellow) are measured by POLQA scores. The centered yellow blob for F-JNF (right plot) corresponds to the position of target speech sources in the dataset. A speech source positioned at the origin represents a speaker that holds the recording tablet frontally at face level. Most speakers in the dataset however tilt the tablet to look at it a bit from above corresponding to a negative latitude value. The yellow blob at the left and right bottom edge shows that the filter cannot differentiate between signals impinging on the microphones attached to the tablet from front-side or back-side, which is expected for a planar microphone array. As the T-JNF has only little spatial selectivity but nevertheless obtains better performance than F-JNF, we conclude that temporal information, which is not reflected in this plot, plays an important role. However, based on the first spatial selectivity plot for FT-JNF, we find that this information can be incorporated without sacrificing a lot of the spatial selectivity, which gives a great performance boost of 0.23 POLQA score. 

In addition, we draw two more general conclusions from the above analysis: First, the plots show clearly that the CHiME3 dataset resembles a scenario with a fixed (only slightly variable) target speaker position relative to the microphone array orientation. This is easily forgotten as the target speaker positions in the CHiME3 dataset are unknown. And second, we have seen that performance improvements observed for a joint multi-channel filter evaluated on the CHiME3 dataset can not directly be attributed to an improved spatial filtering, but that a much more detailed analysis is necessary to understand the internal functioning of such a filter.

Finally, we compare our proposed algorithm with the four additional baselines described in Section \ref{subsec:baselines}. The results are presented in the bottom part of Table \ref{table:chime-results}. The results are consistent with the performances reported on the speaker extraction dataset. Only T-JNF \cite{li2019narrowband} improves in comparison with the other baselines and now slightly outperforms EaBNet \cite{li2022eabnet}. Overall, we find that our proposed architecture FT-JNF, which has been designed to use all three sources of information, outperforms all other baselines regardless of the noise type.

\section{Conclusion}
In this work, we have presented a detailed analysis of the internal mechanisms of a DNN-based filter for multi-channel speech enhancement. While traditional approaches combine a linear spatial filter with a separate tempo-spectral post-filter, DNN-based filters can potentially overcome the linear processing model and exploit interdependencies between spatial and tempo-spectral information. Here, we have shown that a non-linear spatial filter indeed outperforms an oracle MVDR on a challenging speaker extraction task with a low number of microphones. Furthermore, our analyses reveal that the interdependencies between spatial and spectral information can successfully be exploited by a DNN-based filter showing that additional spectral information increases the spatial selectivity of the filter. Our systematic review of this interplay of spatial and and tempo-spectral information leads to a simple network architecture with only two LSTM layers and a single feed-forward layer, that outperforms state-of-the-art network architectures for multi-channel speech enhancement by at least 0.22 POLQA score on the speaker extraction task and 0.32 POLQA score on the CHiME3 noise data. 

\section*{Acknowledgment}
We would like to thank J. Berger and Rohde\&Schwarz SwissQual AG for their support with POLQA.

\bibliographystyle{IEEEtran}
\bibliography{bib}

\begin{IEEEbiography}[{\includegraphics[width=1in,height=1.25in,clip,keepaspectratio]{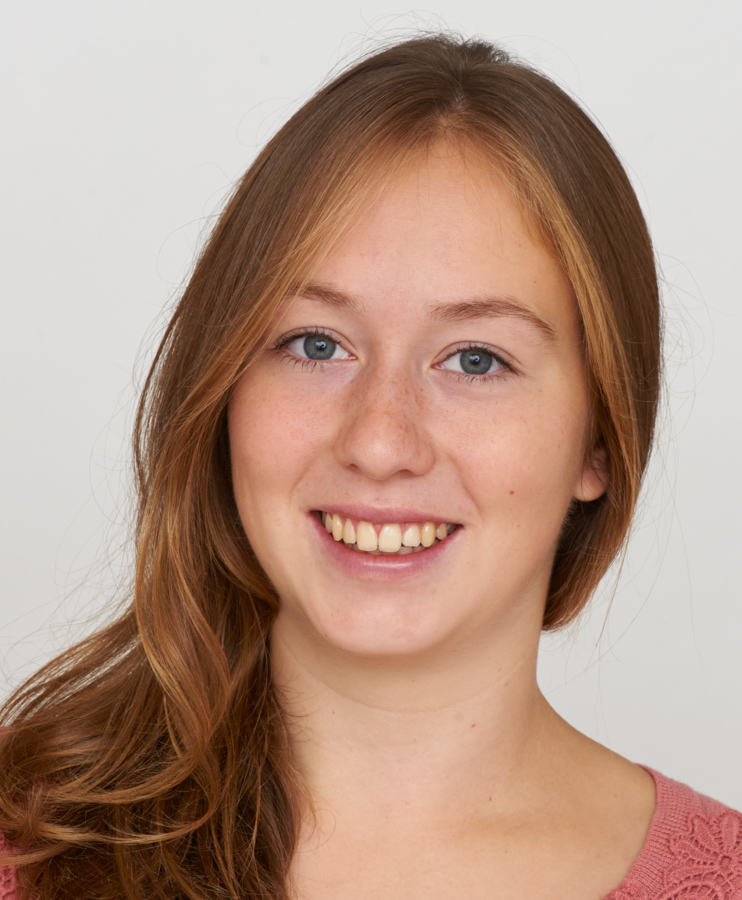}}]{Kristina Tesch}
(S’20) received a B.Sc and M.Sc in Informatics in 2016 and 2019 from the Universität Hamburg, Hamburg, Germany. She is with the Signal Processing Group at the Universität Hamburg since 2019 and is currently working towards a doctoral degree. Her research interests include digital signal processing and machine learning algorithms for speech and audio with a focus on multichannel speech enhancement. %
For her master thesis on multichannel speech enhancement, she received the award for the best master thesis at a German Informatics Department from the Fakultätentag Informatik in 2019, and her prior work on nonlinear spatial filtering \cite{tesch2021nonlinearspatialfilteringtasl} received the VDE ITG 2022 award.
\end{IEEEbiography}

\begin{IEEEbiography}[{\includegraphics[width=1in,height=1.25in,clip,keepaspectratio]{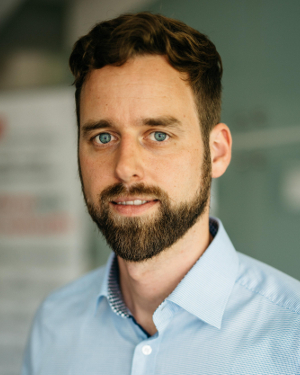}}]{Timo Gerkmann}
(S’08–M’10–SM’15) studied Electrical Engineering and Information Sciences at the Universität Bremen and the Ruhr-Universität Bochum in Germany. He received his Dipl.-Ing. degree in 2004 and his Dr.-Ing. degree in 2010 both in Electrical Engineering and Information Sciences from the Ruhr-Universität Bochum, Bochum, Germany. In 2005, he spent six months with Siemens Corporate Research in Princeton, NJ, USA. During 2010 to 2011 Dr. Gerkmann was a postdoctoral researcher at the Sound and Image Processing Lab at the Royal Institute of Technology (KTH), Stockholm, Sweden. From 2011 to 2015 he was a professor for Speech Signal Processing at the Universität Oldenburg, Oldenburg, Germany. During 2015 to 2016 he was a Principal Scientist for Audio \& Acoustics at Technicolor Research \& Innovation in Hanover, Germany. Since 2016 he is a professor for Signal Processing at the Universität Hamburg, Germany. His main research interests are on statistical signal processing and machine learning for speech and audio applied to communication devices, hearing instruments, audio-visual media, and human-machine interfaces. Timo Gerkmann serves as an elected member of the IEEE Signal Processing Society Technical Committee on Audio and Acoustic Signal Processing and as an Associate Editor of the IEEE/ACM Transactions on Audio, Speech and Language Processing. His prior work on nonlinear spatial filtering \cite{tesch2021nonlinearspatialfilteringtasl} received the VDE ITG 2022 award.
\end{IEEEbiography}

\end{document}